\documentclass{ws-procs961x669}            
\begin{document}

\newcommand{\rma}{\rho_m}
\newcommand{\rr}{\rho_r}
\newcommand{\rL}{\rho_\Lambda}
\newcommand{\PL}{P_\Lambda}
\newcommand{\dG}{\dot{G}}
\newcommand{\drr}{\dot{\rho}_r}
\newcommand{\drL}{\dot{\rho}_\Lambda}
\newcommand{\dH}{\dot{H}}
\newcommand{\CC}{\Lambda}

\newcommand{\rv}{\rho_{\rm vac}}
\newcommand{\Pv}{P_{\rm vac}}
\newcommand{\rvo}{\rho^0_{\rm vac}}

\newcommand{\OM}{\Omega_M}
\newcommand{\Om}{\Omega_m}
\newcommand{\Omo}{\Omega^0_{m}}
\newcommand{\Omh}{\hat{\Omega}_m}
\newcommand{\OMo}{\Omega_{M}^0}
\newcommand{\ORo}{\Omega_{r}}
\newcommand{\OL}{\Omega_{\Lambda}}
\newcommand{\Oc}{\Omega_{c}}
\newcommand{\Oco}{\Omega_{c 0}}
\newcommand{\OLh}{\hat{\Omega}_{\Lambda}}
\newcommand{\GL}{\Gamma_{\Lambda}}
\newcommand{\GLh}{\hat{\Gamma}_{\Lambda}}
\newcommand{\OLo}{\Omega^0_{\Lambda}}
\newcommand{\OX}{\Omega_{X}}
\newcommand{\OXo}{\Omega_{X}^0}
\newcommand{\OXh}{\hat{\Omega}_{X}}
\newcommand{\OD}{\Omega_{D}}
\newcommand{\ODo}{\Omega_{D}}
\newcommand{\OR}{\Omega_R}
\newcommand{\OK}{\Omega_K}
\newcommand{\OKo}{\Omega_{K}^0}
\newcommand{\OZ}{\Omega_0}
\newcommand{\OT}{\Omega_T}
\newcommand{\rc}{\rho_c}
\newcommand{\rco}{\rho^0_{c}}
\newcommand{\rmo}{\rho_{m 0}}
\newcommand{\rs}{\rho_s}
\newcommand{\ps}{p_s}
\newcommand{\rM}{\rho_M}
\newcommand{\rmr}{\rho_m}
\newcommand{\pmr}{p_m}
\newcommand{\rMo}{\rho_{M}^0}
\newcommand{\pM}{p_M}
\newcommand{\rR}{\rho_r}
\newcommand{\rD}{\rho_D}
\newcommand{\rDt}{\tilde{\rho}_D}
\newcommand{\rDo}{\rho_{D}^0}
\newcommand{\rX}{\rho_X}
\newcommand{\pX}{p_X}
\newcommand{\wX}{\omega_X}
\newcommand{\wm}{\omega_m}
\newcommand{\wR}{\omega_R}
\newcommand{\aR}{\alpha_R}
\newcommand{\amr}{\alpha_m}
\newcommand{\aef}{\alpha_e}
\newcommand{\aX}{\alpha_X}
\newcommand{\rLo}{\rho^0_{\CC}}
\newcommand{\pD}{p_D}
\newcommand{\wD}{\omega_D}
\newcommand{\zm}{z_{\rm max}}
\newcommand{\wL}{\omega_{\CC}}
\newcommand{\CCo}{\Lambda_0}
\newcommand{\we}{\omega_{e}}
\newcommand{\re}{r_{\epsilon}}
\newcommand{\tOM}{\tilde{\Omega}_M}
\newcommand{\tOm}{\tilde{\Omega}_m}
\newcommand{\tOmo}{\tilde{\Omega}_m^0}
\newcommand{\tOL}{\tilde{\Omega}_{\CC}}
\newcommand{\tOD}{\tilde{\Omega}_{D}}
\newcommand{\tODo}{\tilde{\Omega}_{D}^0}
\newcommand{\xL}{\xi_{\CC}}
\newcommand{\fM}{f_{M}}
\newcommand{\fL}{f_{\Lambda}}
\newcommand{\lu}{\lambda_1}
\newcommand{\ld}{\lambda_2}
\newcommand{\lt}{\lambda_3}
\newcommand{\model}{X$\CC$CDM}
\newcommand{\f}{\tilde{f}}
\newcommand{\cM}{{\cal M}}
\newcommand{\cMd}{{\cal M}^2}
\newcommand{\ka}{\kappa}

\newcommand{\bCC}{\bar{\CC}}
\newcommand{\bCDM}{\bar{\CC}{\rm CDM}}

\newcommand{\CH}{C_H}
\newcommand{\CHd}{C_{\dot{H}}}
\newcommand{\rDE}{\rho_{\rm DE}}
\newcommand{\tetm}{\theta_{\rm m}}
\newcommand{\rplu}{r_{+}}
\newcommand{\rmin}{r_{-}}
\newcommand{\nueff}{\nu_{\rm eff}}
\newcommand{\nueffp}{\nu_{\rm eff}'}
\newcommand{\xim}{\xi_m}
\newcommand{\xiR}{\xi'}
\newcommand{\rRo}{\rho_{r 0}}
\newcommand{\bk}{{\bf k}}
\newcommand{\mpl}{m_{\rm Pl}}
\newcommand{\MPl}{{\cal M}_{\rm Pl}}

\newcommand{\be}{\begin{equation}}
\newcommand{\ee}{\end{equation}}


\newcommand{\cH}{\mathcal{H}}
\newcommand{\cpH}{\mathcal{H}^\prime}
\newcommand{\cpHs}{\mathcal{H}^{\prime 2}}
\newcommand{\cppH}{\mathcal{H}^{\prime \prime}}
\newcommand{\txi}{\tilde{\xi}}
\newcommand{\ha}{\hat{a}}
\newcommand{\astar}{a_{*}}
\newcommand{\trI}{\tilde{\rho}_I}
\newcommand{\rI}{\rho_I}
\newcommand{\TI}{T_I}
\newcommand{\tHI}{\tilde{H}_I}

\hyphenation{theo-re-ti-cal gra-vi-ta-tio-nal theo-re-ti-cally}


\title{Running vacuum interacting with dark matter or with running gravitational coupling. Phenomenological implications}

\author{Joan Sol\`a Peracaula$^{a,b}$}

\address{$^a$Departament de F\'isica Qu\`antica i Astrof\'isica, \\
and \\ $^b$ Institute of Cosmos Sciences,\\ Universitat de Barcelona, \\
Av. Diagonal 647, E-08028 Barcelona, Catalonia, Spain,
\vspace{0.1cm}\\
$^*$E-mail: sola@fqa.ub.edu}

\begin{abstract}
The cosmological term, $\Lambda$, in Einstein's equations is an essential ingredient of the `concordance' $\Lambda$CDM model of cosmology. In this mini-review presentation, we assess the possibility that $\Lambda$ can be a dynamical quantity, more specifically a `running quantity' in quantum field theory in curved spacetime. A great deal of phenomenological works have shown in the last few years that this option (sometimes accompanied with a running gravitational coupling) may cure some of the tensions afflicting the $\Lambda$CDM. The `running vacuum models' (RVM's) are characterized by the vacuum energy density, $\rho_{\rm vac}$, being a series of (even) powers of the Hubble rate and its time derivatives. Here we describe the technical quantum field theoretical origin of the RVM structure in FLRW spacetime, which goes well-beyond the original semi-qualitative renormalization group arguments. In particular, we compute the renormalized energy-momentum tensor using the adiabatic regularization procedure and show that it leads to the RVM form. In other words, we find that the renormalized vacuum energy density, $\rho_{\rm vac}(H)$ evolves as a (constant) additive term plus leading dynamical components ${\cal O}(H^2)$. There are also ${\cal O}(H^4)$ contributions, which can be relevant for the early universe. Remarkably enough, the renormalized  $\rho_{\rm vac}(H)$ does not exhibit dangerous terms proportional to the quartic power of the masses ($\sim m^4$) of the fields. It is well-known that these terms have been the main source of trouble since they are responsible for the extreme fine tuning and ultimately for the cosmological constant problem. In its canonical form, the current $\rho_{\rm vac}(H)$ is dominated by a constant term, as it should be, but it acquires a mild dynamical component $\sim \nu H^2$ ($0<\nu\ll1$) which makes the RVM to mimic quintessence.
\end{abstract}

\keywords{Running Vacuum,  Dynamical Dark Energy, Inflation, Dark Matter}

\bodymatter

\section{Introduction}\label{sec:intro}

The standard or concordance model of cosmology, the so-called $\CC$CDM,  is a very successful  theoretical framework for the description of the Universe\cite{Peebles1984}.   Crucial ingredients of it are left, though,  without direct observational evidence and/or a proper theoretical interpretation  based on fundamental principles.  Such is the situation with dark matter, but also with the cosmological term, $\CC$, which has been traditionally associated with the vacuum energy density (VED) of the Universe, $\rv$.  In fact, the  notion of VED in cosmology is a most subtle concept, which is challenging theoretical physicists and cosmologists for many decades, specially with the advent of Quantum Theory in general and the more sophisticated machinery of  Quantum Field Theory (QFT). The roots of the problem reside in the interpretation of the cosmological constant (CC) term,  $\CC$,  in Einstein's equations as a quantity being connected with the VED, which is also a fundamental concept in QFT.    The proposed connection is $\rv=\CC/(8\pi G_N)$,  where $G_N$ is Newton's constant.  Accurate measurements of $\rv$ in the last decades from distant type Ia supernovae (SnIa)  and the cosmic microwave background (CMB) \cite{SNIaOrig,SNIa2018,PlanckCollab}, have put the foundations of the concordance  $\CC$CDM model of cosmology\cite{Peebles1984}.

The concordance model  is formulated in  the context of the Friedman-Lemaitre-Robertson-Walker (FLRW) framework and is deeply ingrained in the General Relativity (GR) paradigm.  However, one of the most important drawbacks of GR  is that it is a non-renormalizable theory.  This can be considered a serious theoretical obstruction for GR to be considered a fundamental theory of gravity. This fact  adversely impacts on the $\CC$CDM too.  GR cannot properly describe the short distance effects of gravity, i.e.  the ultraviolet regime (UV), only the large distance effects (or infrared regime).  As a consequence,  GR cannot provide by itself a framework for quantizing gravity (the spacetime metric field)  along with the rest of the fundamental  interactions  (assuming of course that gravity is amenable to be  quantized on conventional grounds). In this sense, a first (rougher but effective) approach is to treat gravity as a classical (external or background) field and quantize only the matter fields of the fundamental interactions (electroweak and strong interactions). This is the main program of the semiclassical approach, namely  the point of view of QFT in curved spacetime, see\cite{BirrellDavies82,ParkerToms09,Fulling89,MukhanovWinitzki07} for a review. This will be our pursue here as well.  We shall nevertheless still be able to compute the quantum effects from matter and their contribution to the VED.  Not only so, it will allow us to renormalize these (originally UV-divergent)  quantum effects and obtain a finite quantity that can be better compared with observations.  The renormalized VED found in the present framework  is free from the traditional $\sim m^4$ effects (proportional to the quartic powers of the masses of the matter fields).  This was first shown in the literature in the  work\cite{CristianJoan2020} and further extended in\cite{CristianJoan2021}.   It was demonstrated that $\rv$   appears (in the current universe)  as a constant (dominant) term plus a dynamical component which varies as $\sim \nu H^2\mpl^2$, with $\nu$  a small (dimensionless) and QFT-computable coefficient, $\mpl$ being the usual Planck mass ($\mpl=G_N^{-1/2}$).  Quite obviously, such a result is highly compatible with the $\CC$CDM since it involves only a small departure from the rigid vacuum term enforced in it. Notwithstanding,  it makes a significant qualitative (and quantitative)  prediction:  the quantum vacuum  does not remain static throughout the cosmic  evolution but is mildly dynamical, and such dynamics can be effectively evaluated from first principles, for  $\nu$ (as mentioned) can be accounted for using QFT in curved spacetime. As long as the vacuum turns out to be free from the weird $\sim m^4$ effects obtained in simplified renormalization treatments, the finite result which is proposed here appears theoretically distinct and  very tantalizing. It suggests to reassess  if the Cosmological Constant Problem (CCP)\cite{Weinberg89,Sahni2000,PeeblesRatra2003,Padmanabhan2003,Copeland2006,DEBook}  is still in force in  the current theoretical framework.  Remember that the CCP  is one of the hardest and longstanding mysteries of theoretical physics,  and proves to be a serious impediment to reconcile cosmology with particle physics and quantum field theory in general\cite{JSPRev2013}.

The smoother formulation of the quantum vacuum as put forward here is not only convenient from the theoretical point of view, but also from  the observational side. It has long been known that, observationally, there appear to be some discrepancies  or ``tensions'' between the results of data analyses from the Planck Collaboration, based on the $\Lambda$CDM~\cite{PlanckCollab}, and the local (distance ladder)  measurements of the Hubble parameter today, the so called $H_0$ tension~\cite{Htension1,Htension2},   see e.g. the various reviews ~\cite{tensions,PerivoSkara2021,ValentinoReviewTensions,TensionsJSP2018}.  Not only so, there exist  also tensions in the large scale structure (LSS) growth data, the so-called $\sigma_8$ tension~\cite{PerivoSkara2021,s8tensionValentino,TensionsJSP2018}.  Although such discordances in the arena of the `concordance' model might admit a variety of more mundane astrophysical explanations~\cite{efst,dust,Biagio}, or even disappear into thin air when  more  data will be available in the near future, it has been conceded already that they may well be passing the point of being attributable to a fluke\cite{Htension2}.

As it turns out, the smooth quantum vacuum phenomenological framework described here has the capacity to deal successfully with  the $H_0$ and the $\sigma_8$ tensions.  Such a framework
represents the running vacuum model of cosmology (RVM) -- see \cite{JSPRev2013} for a detailed review and references therein. The latter  was originally motivated by semi-qualitative renormalization group arguments~\cite{ShapSol}, which were  later on formulated in some cases from an action functional viewpoint\,\cite{Fossil}  --  and was finally substantiated from full-fledged  QFT calculations in\cite{CristianJoan2020,CristianJoan2021}. For related studies of vacuum energy in QFT, see e.g. \cite{Babic2005,Maggiore2011,Bilic2011,KohriMatsui2017,FerreiroNavarroSalas,JMarianJentschura2021}.
In this summarized account I will mainly review the emergence of the  RVM from the mentioned  QFT calculations  in curved spacetime, specifically in the context of the FLRW background as well as the most recent phenomenological applications.  It is remarkable that it is also possible to derive a string-inspired version of the RVM cosmological model with  gravitational anomalies in the early Universe, see \cite{EPJ-ST} for a detailed exposition and ~\cite{BasMavSol,ms1,NickPhiloTrans,NickMG16} for more details and spin-off possibilities.

The structure of our presentation is as follows. In sections 2 and 3 we define our basic QFT framework in curved spacetime and the adiabatic expansion procedure. In section\,\ref{sec:RenormEMT} we renormalize the energy-momentum tensor  in the FLRW context and make contact with the RVM.   In section \ref{sec:PhenoApplications} we apply the RVM to tackle the $\sigma_8$ and $H_0$ tensions.    Finally, section \ref{sec:conclusions} contains our conclusions and outlook.

\section{Non-minimally coupled scalar field in FLRW background}\label{sec:EMT}

We consider  the semiclassical calculation of the energy density of the vacuum fluctuations of a quantized scalar field in FLRW spacetime.  It serves as a representative  case of study for the kind of results and difficulties we expect to encounter when dealing with generic quantized fields in a curved background.   Such QFT calculation  implies to perform renormalization since we meet UV-divergent integrals. Conventional approaches such as e.g. the minimal subtraction scheme  lead to  quartic dependence on the mass of the field ($\sim m^4$) enforcing a serious fine tuning among the parameters,  see e.g. \cite{JSPRev2013} and references therein.  Here we avoid using such an unsuccessful method and adopt the adiabatic renormalization procedure (ARP)\cite{BirrellDavies82,ParkerToms09,Fulling89}.  However, we follow the specific approach  presented in\cite{CristianJoan2020}, which is related to that of \cite{FerreiroNavarroSalas}.  For a full-fledged exposition and calculation details,  see \cite{CristianJoan2021}.

\subsection{Action and classical energy-momentum tensor}\label{Sec:FieldEquations}

We start from the  Einstein-Hilbert (EH) action for gravity plus matter\footnote{Our conventions and other useful formulae  are those of \cite{CristianJoan2020,CristianJoan2021}, see the appendices of these references.}:
\begin{equation}\label{eq:EH}
S_{\rm EH+m}=  \frac{1}{16\pi G_N}\int d^4 x \sqrt{-g}\, R  -  \int d^4 x \sqrt{-g}\, \rL+ S_{\rm m}\,.
\end{equation}
The matter action $ S_{\rm m}$ is generic at this point, but will be specified shortly. The (constant) term $\rL$ has dimension of energy density.  Because it is unrelated to the matter part (fully contained in $ S_{\rm m}$), $\rL$ is usually called the vacuum energy density (VED). However, we will not call it that way here since it is not yet the physical VED, $\rv$,  as we shall see.  For us the term $\rL$ is  just a bare parameter of the EH action, as the gravitational coupling  $G_N$ itself. The physical values can only be identified   after renormalizing the bare theory. The corresponding  gravitational field equations emerging from the variation of the action  \eqref{eq:EH} can be put in the convenient form
\begin{equation} \label{FieldEq2}
\MPl^2\, G_{\mu \nu}=-\rho_\Lambda g_{\mu \nu}+T_{\mu \nu}^{\rm m}\,,
\end{equation}
where $\MPl=\mpl/\sqrt{8\pi}=1/\sqrt{8\pi G_N}$ is the reduced Planck mass, the expression $G_{\mu\nu}=R_{\mu\nu}-(1/2) g_{\mu\nu} R$   is the usual Einstein tensor and  $ T_{\mu \nu}^{\rm m}$  is the  stress-energy-momentum tensor, or just energy-momentum tensor (EMT for short) of matter:
\begin{equation}\label{eq:deltaTmunu2}
  T_{\mu \nu}^{\rm m}=-\frac{2}{\sqrt{-g}}\frac{\delta S_{\rm m}}{\delta g^{\mu\nu}}\,.
\end{equation}
 For simplicity we will assume that there is only one (matter) quantum  field contribution to the EMT  on the right hand side of \eqref{FieldEq2} in the form of  a real scalar field, $\phi$. Such contribution will be denoted $T_{\mu \nu}^{\phi}$.   We neglect for the moment the incoherent matter contributions  from dust and radiation.  They can be added a posteriori  without altering the pure  QFT aspects on  which we wish to focus at this point.  We assume that $\phi$ is non-minimally coupled to gravity. Thus, the part of the action involving $\phi$ reads
\begin{equation}\label{eq:Sphi}
  S[\phi]=-\int d^4x \sqrt{-g}\left(\frac{1}{2}g^{\mu \nu}\partial_{\mu} \phi \partial_{\nu} \phi+\frac{1}{2}(m^2+\xi R)\phi^2 \right)\,.
\end{equation}
The non-minimal coupling of $\phi$ to gravity is  $\xi$.  For $\xi=1/6$, the massless ($m=0$)  action is conformally invariant.
We will keep $\xi$ general to explore its influence.   We  assume also that $\phi$   does not couple to itself and hence we shall not consider  a possible contribution from a classical potential for $\phi$ in our analysis.  In this study, we wish to target mainly the zero-point energy (ZPE)  of $\phi$.

The classical  EMT can be derived from the action \eqref{eq:Sphi} and reads as follows:
\begin{equation}\label{EMTScalarField}
\begin{split}
T_{\mu \nu}^{\phi}=&-\frac{2}{\sqrt{-g}}\frac{\delta S[\phi]}{\delta g^{\mu\nu}}= (1-2\xi) \partial_\mu \phi \partial_\nu\phi+\left(2\xi-\frac{1}{2} \right)g_{\mu \nu}\partial^\sigma \phi \partial_\sigma\phi\\
& -2\xi \phi \nabla_\mu \nabla_\nu \phi+2\xi g_{\mu \nu }\phi \Box \phi +\xi G_{\mu \nu}\phi^2-\frac{1}{2}m^2 g_{\mu \nu} \phi^2.
\end{split}
\end{equation}
 Varying the action \eqref{eq:Sphi} with respect to $\phi$ we find the  Klein-Gordon (KG) equation in curved spacetime:
\be\label{eq:KG}
(\Box-m^2-\xi R)\phi=0\,,
\ee
where $\Box\phi=g^{\mu\nu}\nabla_\mu\nabla_\nu\phi=(-g)^{-1/2}\partial_\mu\left(\sqrt{-g}\, g^{\mu\nu}\partial_\nu\phi\right)$. The FLRW line element for spatially flat three-dimensional geometry can be written in conformal coordinates as  $ds^2=a^2(\tau)\eta_{\mu\nu}dx^\mu dx^\nu$, where  $\eta_{\mu\nu}={\rm diag} (-1, +1, +1, +1)$ is the Minkowski  metric in our conventions.   The derivative with respect to the conformal time, $\tau$, will be denoted   $^\prime\equiv d/d\tau$ and thus the  Hubble rate in conformal time  reads $\mathcal{H}(\tau)\equiv a^\prime /a$.  Since  $dt=a d\tau$,  the relation between the Hubble rate in cosmic and conformal times is simply $\mathcal{H}(\tau)=a  H(t)$, with  $H(t)=\dot{a}/a$ ( $\dot{}\equiv d/dt$)  the usual Hubble rate.

 The KG equation \eqref{eq:KG} in conformally flat coordinates becomes
\begin{equation}\label{eq:KGexplicit}
 \phi''+2\cH\phi'-\nabla^2\phi+a^2(m^2+\xi R)\phi=0\,,
\end{equation}
where we used the curvature scalar  of spacetime:  $R=6a^{\prime\prime}/a^3$ .
The separation of variables  in these coordinates, namely $\phi(\tau,x)\sim \int d^3k \ A_{\bf k}\psi_k({\bf x})\phi_k(\tau)+cc$, can  be achieved with $\psi_k(x)=e^{i{\bf k\cdot x}}$. However, in contrast to the Minkowski case we cannot take  $\phi_k(\tau)=e^{\pm i\omega_k \tau}$ since the mode frequencies are not constant anymore.  The form of the  modes $\phi_k(\tau)$  in the curved spacetime case are determined by the KG equation.  Starting from the Fourier expansion with separated space and time variables
\begin{equation}\label{FourierModes}
\phi(\tau,{\bf x})=\int\frac{d^3{k}}{(2\pi)^{3/2}} \left[ A_\bk e^{i{\bf k\cdot x}} \phi_k(\tau)+A_\bk^\ast e^{-i{\bf k\cdot x}} \phi_k^*(\tau) \right]
\end{equation}
(in which  $A_\bk $  and their complex conjugates $A_\bk^\ast$  are the classical Fourier coefficients) and substituting it into \eqref{eq:KGexplicit} the mode functions $\phi_k(\tau)$ are determined by solving the nontrivial differential equation
\begin{equation}\label{eq:KGFourier}
 \phi_k''+2\cH\phi'_k+\left(\omega_k^2(m)+a^2\xi R\right)\phi_k=0\,,
\end{equation}
where $\omega_k^2(m)\equiv k^2+a^2 m^2$.  The mode functions depend only on the modulus $k\equiv|\bk|$ of the comoving momenta,  being $\tilde{k}=k/a$ the physical ones. The frequencies are in general functions of the time-evolving scale factor $a=a(\tau)$, and this makes the particle interpretation hard. If we perform the change of field mode variable  $\phi_k=\varphi_k/a$   the above equation simplifies to
\begin{equation}\label{eq:KGFourier2}
\varphi_k^{\prime \prime}+\left(\omega_k^2(m)+a^2\,(\xi-1/6)R)\right)\varphi_k=0\,.
\end{equation}
For conformally invariant matter ($m=0$ and $\xi=1/6$),  such equation boils down to the form $\varphi_k^{\prime \prime}+k^2\varphi_k=0$, whose positive- and negative-energy solutions are just  $e^{- ik\tau}$  and $e^{+ ik\tau}$, respectively.  On the other hand, in the massless case with minimal coupling ($\xi=0$)  the previous equation further simplifies to
$
\varphi_k''+(k^2-a^2R/6)\varphi_k=0\,.
$
In the radiation epoch ($a\propto\tau$, thus  $R=6a^{\prime\prime}/a^3=0$) we find once more  the trivial modes $\varphi_k(\tau)=e^{\pm ik\tau}$.   Both in the de Sitter ($a=-1/(H\tau)$,  $H=$const.) and matter-dominated ($a\propto\tau^2$) epochs we have $a^2R=12/\tau^2$, which leads to
$\varphi_k''+(k^2-2/\tau^2)\varphi_k=0\,.$
This equation  admits an exact (positive-energy) solution in terms of  Hankel  functions of half integer order, hence in close analytic form. In the de Sitter case ($\tau<0$) one may impose the Bunch-Davies vacuum limit $\sim e^{-ik|\tau|}$ in the far remote past ($\tau\to-\infty$) and  one finds $\varphi(\tau)\propto\ (1-i/(k|\tau|))e^{-ik|\tau|}$. The same solution is valid for the matter-dominated era (for which $\tau>0$).     If, however,  $m\neq0$ and/or $\xi\neq 1/6$ no analytic solution of \eqref{eq:KGFourier} is available, and this leads us to perform  a WKB (Wentzel-Kramers-Brillouin)  expansion of the solution.  But before tackling that method, let us consider the quantization of the scalar field $\phi$.

 \subsection{Quantum fluctuations}\label{sec:AdiabaticVacuum}

To account for the quantum fluctuations of the scalar field $\phi$  we must address the expansion of the field around its background value $\phi_b$:
\begin{equation}
\phi(\tau,x)=\phi_b(\tau)+\delta\phi (\tau,x). \label{ExpansionField}
\end{equation}
One starts defining  an appropriate vacuum state, called adiabatic vacuum\cite{Bunch1980}. The  vacuum expectation value (VEV) of $\phi$  is  identified with the background value,  $\langle 0 | \phi (\tau, x) | 0\rangle=\phi_b (\tau)$, whereas the VEV of the fluctuation is zero:  $\langle  \delta\phi  \rangle\equiv \langle 0 | \delta\phi | 0\rangle =0$. This is not the case for the VEV of the bilinear products of fluctuations, e.g. $\langle \delta\phi^2 \rangle\neq0$.  It is convenient to decompose  $\langle T_{\mu \nu}^\phi \rangle=\langle T_{\mu \nu}^{\phi_b} \rangle+\langle T_{\mu \nu}^{\delta\phi}\rangle$, where
$\langle T_{\mu \nu}^{\phi_{b}} \rangle =T_{\mu \nu}^{\phi_{b}} $
is the  contribution  from the classical background part,  whereas $\langle T_{\mu \nu}^{\delta\phi}\rangle\equiv \langle 0 | T_{\mu \nu}^{\delta\phi}| 0\rangle$  is  the genuine vacuum contribution from the field fluctuations.  Taking into account that  $\rho_\Lambda$  is also part of the vacuum action  \eqref{eq:EH}, the full vacuum contribution is  the sum
\begin{equation}\label{EMTvacuum}
\langle T_{\mu \nu}^{\rm vac} \rangle=-\rho_\Lambda g_{\mu \nu}+\langle T_{\mu \nu}^{\delta \phi}\rangle\,.
\end{equation}
Thus, the total vacuum part  receives contributions from both the cosmological term in the action as well as from the quantum fluctuations  of the field (the ZPE). However, since these quantities are formally UV-divergent, the physical vacuum can only be identified a posteriori, namely upon suitable regularization and renormalization. For this we adopt the adiabatic method along the lines of\cite{CristianJoan2020,CristianJoan2021}.

Obviously the classical and  quantum parts of the  field (\ref{ExpansionField}) obey the  curved spacetime KG equation  \eqref{eq:KGFourier} separately.  Similarly for $\varphi=\varphi_b+\delta\varphi$  (where $\phi=\varphi/a$).  Denoting the frequency modes of the fluctuating part  $\delta\varphi$ by   $h_k(\tau)$, we can write
\begin{equation}\label{FourierModesFluc}
\delta \varphi(\tau,{\bf x})=\int \frac{d^3{k}}{(2\pi)^{3/2}} \left[ A_\bk e^{i{\bf k\cdot x}} h_k(\tau)+A_\bk^\dagger e^{-i{\bf k\cdot x}} h_k^*(\tau) \right]\,.
\end{equation}
Here  $A_\bk$ and  $A_\bk^\dagger $ are now the (time-independent) annihilation and creation operators, which satisfy the commutation relations
\begin{equation}\label{CommutationRelation}
[A_\bk, A_\bk'^\dagger]=\delta({\bf k}-{\bf k'}), \qquad [A_\bk,A_ \bk']=0.
\end{equation}
The frequency modes of the fluctuations, $h_k(\tau)$, satisfy the  differential equation
\begin{equation}\label{eq:ODEmodefunctions}
h_k^{\prime \prime}+\Omega_k^2(\tau) h_k=0\ \ \ \ \ \ \ \ \ \ \ \Omega_k^2(\tau) \equiv\omega_k^2(m)+a^2\, (\xi-1/6)R\,.
\end{equation}
Except in the simple cases mentioned above, the solution of that equation requires a recursive self-consistent iteration,  the  WKB expansion.  One start from
\begin{equation}\label{eq:phaseIntegral}
h_k(\tau)=\frac{1}{\sqrt{2W_k(\tau)}}\exp\left(i\int^\tau W_k(\tilde{\tau})d\tilde{\tau} \right)\,,
\end{equation}
where the  normalization factor $1/\sqrt{2W_k(\tau)}$ insures that the Wronskian condition
$ h_k^\prime h_k^* -  h_k^{} h_k^{*\prime}=i$ 
is satisfied. It warrants the standard equal-time commutation relations.
Functions $W_k$ in the above ansatz obey the (non-linear) equation
\begin{equation} \label{WKBIteration}
W_k^2(\tau)=\Omega_k^2(\tau) -\frac{1}{2}\frac{W_k^{\prime \prime}}{W_k}+\frac{3}{4}\left( \frac{W_k^\prime}{W_k}\right)^2\,,
\end{equation}
which is amenable to be  solved using the WKB expansion. The latter is applicable only for large $k$, therefore  short wave lengths (as e.g. in geometrical Optics),  and weak gravitational fields.  The mode functions $h_k(\tau)$ are no longer of the form  $\varphi_k(\tau)=e^{\pm i\omega_k\tau}$,  hence particles with definite frequencies cannot be strictly defined in a curved background.  Notwithstanding,  an approximate Fock space interpretation is still feasible if the  vacuum is defined as the quantum state which is annihilated by all the operators $A_{\bf k}$ of the above Fourier expansion. This defines in a precise way the notion of  the adiabatic vacuum\cite{Bunch1980,BirrellDavies82,ParkerToms09,Fulling89,MukhanovWinitzki07}.

\subsection{WKB expansion of the mode functions}\label{sec:WKB}

 In the gravitational context, the  WKB expansion leads to the adiabatic regularization procedure (ARP).  For a review, see e.g.  the classic books \cite{BirrellDavies82,ParkerToms09}.  The regularization involved in the ARP amounts to subtracted integrals which become UV-finite and hence one obtains  direct renormalization of the physical quantities.  In the two successives works\cite{CristianJoan2020,CristianJoan2021} the WKB expansion was performed first up to $4th$ and subsequently up to $6th$ adiabatic order. In the latter case the calculational details are rather cumbersome, but are necessary in order to study the on-shell renormalized theory. Here, however, we will limit ourselves to describe the results up to $4th$ order, which is enough to renormalize the theory off-shell.   The counting of adiabatic orders in the WKB expansion follows the number of time derivatives. Thus:  $k^2$ and $a$ are of adiabatic order $0$;  $a^\prime$ and $\mathcal{H}$  of adiabatic order 1;  $a^{\prime \prime},a^{\prime 2},\mathcal{H}^\prime$ and $\mathcal{H}^2$ as well as $R$ are of adiabatic order $2$. Each additional derivative increases the adiabatic order  by one unit.   The expansion collects the different adiabatic orders:
\begin{equation}\label{WKB}
W_k=\omega_k^{(0)}+\omega_k^{(2)}+\omega_k^{(4)}+\omega_k^{(6)}\cdots,
\end{equation}
General covariance precludes the odd adiabatic orders.  The  $\omega_k^{(j)}$ can be expressed in terms of $\Omega_k(\tau)$ and its time derivatives.
Following \cite{CristianJoan2020,CristianJoan2021} we consider an off-shell procedure  in which the frequency $\omega_k$ of a given mode  is defined not at the mass $m$ of the particle but at an arbitrary mass scale $M$:
\be\label{eq:omegaM}
\omega_k\equiv\omega_k(\tau, M)\equiv \sqrt{k^2+a^2(\tau) M^2}\,.
\ee
At the moment we will use just the notation $\omega_k$ to indicate such off-shell value, and when necessary we will distinguish it from the on-shell one using the forms $\omega_k(M)$ and  $\omega_k(m)$, both being of course functions of $\tau$ (which we will omit to simplify notation).
Working out  the second and  fourth order terms of \eqref{WKB} one finds\cite{CristianJoan2020}
\begin{equation}
\begin{split}
\omega_k^{(0)}&= \omega_k\,,\\
\omega_k^{(2)}&= \frac{a^2 \Delta^2}{2\omega_k}+\frac{a^2 R}{2\omega_k}(\xi-1/6)-\frac{\omega_k^{\prime \prime}}{4\omega_k^2}+\frac{3\omega_k^{\prime 2}}{8\omega_k^3}\,,\\
\omega_k^{(4)}&=-\frac{1}{2\omega_k}\left(\omega_k^{(2)}\right)^2+\frac{\omega_k^{(2)}\omega_k^{\prime \prime}}{4\omega_k^3}-\frac{\omega_k^{(2)\prime\prime}}{4\omega_k^2}-\frac{3\omega_k^{(2)}\omega_k^{\prime 2}}{4\omega_k^4}+\frac{3\omega_k^\prime \omega_k^{(2)\prime}}{4\omega_k^3}\,.
\end{split}\label{WKBexpansions1}
\end{equation}
The quadratic mass differences  $\Delta^2\equiv m^2-M^2$  must be counted as being of adiabatic order 2 since they appear in the WKB expansion along with other terms of the same adiabatic order\footnote{In the context of the effective action,  it  can be justified more formally on replacing $m$ by $M$ in the heat kernel expansion of the propagator, but we shall not take this path here, see\cite{CristianJoan2021}  for details. }.  The on-shell result is recovered for $M=m$, for which  $\Delta = 0$ and corresponds to the usual ARP procedure\cite{BirrellDavies82,ParkerToms09}.   One could extend the expansion up to the next nonvanishing  adiabatic order, which is order $6th$, although we refrain from quoting the result here\cite{CristianJoan2021}.  It is easy to see that the adiabatic expansion becomes an expansion in powers of $\mathcal{H}$ and its time derivatives.
For example, the  first two derivatives of $\omega_k$  read
\begin{equation}\label{omegak0}
\omega_k^\prime=a^2\mathcal{H}\frac{M^2}{\omega_k}, \qquad\omega_k^{\prime \prime}=2a^2\mathcal{H}^2\frac{M^2}{\omega_k}+a^2\mathcal{H}^\prime \frac{M^2}{\omega_k}-a^4\mathcal{H}^2\frac{M^4}{\omega_k^3}\,,
\end{equation}
where we recall that $\mathcal{H}$ is the Hubble function in conformal time. From these elementary differentiations one can then compute the more laborious derivatives appearing in the above expressions, such as $ \omega_k^{(2)\prime}, \omega_k^{(2)\prime\prime}$ etc.  Therefore, the final result appears as an expansion in powers of $\cH$ and multiple derivatives of it.

\section{Adiabatic expansion of the ZPE }\label{eq:RegZPE}

We have now all the necessary ingredients  to compute the zero-point energy (ZPE) associated to the quantum vacuum fluctuations in curved spacetime with FLRW metric.  We closely follow the presentation of \cite{CristianJoan2020}.  Inserting the decomposition (\ref{ExpansionField}) of the quantum field $\phi$ in the  EMT as given in Eq.\,\eqref{EMTScalarField} and selecting only the fluctuating parts  $\delta\phi$, the ZPE  (which is ssociated to the  $00$-component)  reads
\begin{equation}\label{EMTInTermsOfDeltaPhi}
\begin{split}
\langle T_{00}^{\delta \phi}\rangle =&\left\langle \frac{1}{2}\left(\delta\phi^{\prime}\right)^2+\left(\frac{1}{2}-2\xi\right)\left(\nabla\delta \phi\right)^2+6\xi\mathcal{H}\delta \phi \delta \phi^\prime\right.\\
&\left.-2\xi\delta\phi\,\nabla^2\delta\phi+3\xi\mathcal{H}^2\delta\phi^2+\frac{a^2m^2}{2}(\delta\phi)^2 \right\rangle\,.
\end{split}
\end{equation}
Notice that $\delta\phi'$, the fluctuation of the differentiated field (with respect to conformal time), is given by $\delta\phi^{\prime}\equiv\delta\partial_0\phi= \partial_0\delta\phi=(\delta\phi)'$. Next we substitute the Fourier expansion of $\delta\phi=\delta\varphi/a$, as given in \eqref{FourierModesFluc},  into Eq.\,\eqref{EMTInTermsOfDeltaPhi} and use the commutation relations \eqref{CommutationRelation}. At the same time we  symmetrize  the operator field products $\delta\phi \delta\phi^\prime$ with respect to the creation and annihilation operators.  We present the final result in Fourier space, and hence  we integrate  $\int\frac{d^3k}{(2\pi)^3}(...)$ over solid angles\cite{CristianJoan2020}:
\begin{equation}\label{EMTFluctuations}
\begin{split}
\langle T_{00}^{\delta \phi (0-4)} \rangle & =\frac{1}{8\pi^2 a^2}\int dk k^2 \left[ 2\omega_k+\frac{a^4M^4 \mathcal{H}^2}{4\omega_k^5}-\frac{a^4 M^4}{16 \omega_k^7}(2\mathcal{H}^{\prime\prime}\mathcal{H}-\mathcal{H}^{\prime 2}+8 \mathcal{H}^\prime \mathcal{H}^2+4\mathcal{H}^4)\right.\\
&+\frac{7a^6 M^6}{8 \omega_k^9}(\mathcal{H}^\prime \mathcal{H}^2+2\mathcal{H}^4) -\frac{105 a^8 M^8 \mathcal{H}^4}{64 \omega_k^{11}}\\
&+\left(\xi-\frac{1}{6}\right)\left(-\frac{6\mathcal{H}^2}{\omega_k}-\frac{6 a^2 M^2\mathcal{H}^2}{\omega_k^3}+\frac{a^2 M^2}{2\omega_k^5}(6\mathcal{H}^{\prime \prime}\mathcal{H}-3\mathcal{H}^{\prime 2}+12\mathcal{H}^\prime \mathcal{H}^2)\right. \\
& \left. -\frac{a^4 M^4}{8\omega_k^7}(120 \mathcal{H}^\prime \mathcal{H}^2 +210 \mathcal{H}^4)+\frac{105a^6 M^6 \mathcal{H}^4}{4\omega_k^9}\right)\\
&+\left. \left(\xi-\frac{1}{6}\right)^2\left(-\frac{1}{4\omega_k^3}(72\mathcal{H}^{\prime\prime}\mathcal{H}-36\mathcal{H}^{\prime 2}-108\mathcal{H}^4)+\frac{54a^2M^2}{\omega_k^5}(\mathcal{H}^\prime \mathcal{H}^2+\mathcal{H}^4) \right)
\right]\\
&+\frac{1}{8\pi^2 a^2} \int dk k^2 \left[  \frac{a^2\Delta^2}{\omega_k} -\frac{a^4 \Delta^4}{4\omega_k^3}+\frac{a^4 \mathcal{H}^2 M^2 \Delta^2}{2\omega_k^5}-\frac{5}{8}\frac{a^6\mathcal{H}^2 M^4\Delta^2}{\omega_k^7} \right.\\
& \left. +\left( \xi-\frac{1}{6} \right) \left(-\frac{3a^2\Delta^2 \mathcal{H}^2}{\omega_k^3}+\frac{9a^4 M^2 \Delta^2 \mathcal{H}^2}{\omega_k^5}\right)\right]\,.
\end{split}
\end{equation}
As expected, only even powers of $\cal H$ remain in the final result.
The  Minkowskian spacetime result for the on-shell ZPE ($M=m$)  is obtained as a very particular case of the above expression for $a=1$ ($\mathcal{H}=0)$:
\begin{equation}\label{eq:Minkoski}
  \left.\langle T_{00}^{\delta \phi}\rangle\right|_{\rm Minkowski}=\frac{1}{4\pi^2}\int dk k^2 \omega_k =  \int\frac{d^3k}{(2\pi)^3}\,\left(\frac12\,\hbar\,\omega_k\right)\,,
\end{equation}
where $\hbar$ has been  restored for convenience only in the last expression. The result is quartically UV-divergent.   Usual attempts (e.g. through the minimal subtraction scheme) to regularize and renormalize this quantity by e.g.  cancelling the corresponding UV-divergence against the bare $\rL$ term  in the action \eqref{eq:EH}  ends up with the well-known fine-tuning  problem,  which is considered to be the weirdest and toughest aspect of the CCP -- see e.g.\cite{JSPRev2013,Akhmedov2002} and references therein.  We will certainly not proceed in this way  here. We seek (and will find) an alternative way.

\section{Renormalization of the VED  in curved spacetime: the RVM}\label{sec:RenormEMT}

The vacuum energy density  in the expanding universe can be compared with a Casimir device in which the parallel plates slowly move apart (“expand”)\cite{JSPRev2013}.
Although the  total VED cannot be measured, the distinctive  effect associated to the presence of the plates, and then also to their increasing separation with time, it can.  In a similar fashion,  in the cosmological spacetime there is a distinctive nonvanishing spacetime curvature $R$ as compared to Minkowskian spacetime that  is changing with the expansion.  We expect that the measurable VED must be that one which is associated to purely geometric contributions proportional to $R$, $R^2$, $R^{\mu\nu}R_{\mu\nu}$ etc., hence to $H^2$ and $\dot{H}$ (including higher powers of these quantities in the early Universe).

Following\cite{CristianJoan2020, CristianJoan2021},   a subtraction of the VEV of the EMT is carried out at an arbitrary mass scale $M$, playing the role of renormalization point.
Taking into account that the only adiabatic orders that are divergent in the case of the EMT are the  first four ones (in $4$-dimensional spacetime), the subtraction at the scale $M$ is performed only up to the fourth adiabatic order. The on-shell value of the EMT can be computed of course at any order. The terms beyond the $4th$ order are  finite.
The renormalized EMT in this context therefore reads
\begin{eqnarray}\label{EMTRenormalized}
\langle T_{\mu\nu}^{\delta \phi}\rangle_{\rm Ren}(M)&=&\langle T_{\mu\nu}^{\delta \phi}\rangle(m)-\langle T_{\mu\nu}^{\delta \phi}\rangle^{(0-4)}(M)\,.
\end{eqnarray}
Let us apply this procedure to the ZPE part of the EMT, as given by  Eq.\,\eqref{EMTFluctuations}. To ease the presentation of the explicit result,  it proves convenient to recover at least in part  the more explicit notation \eqref{eq:omegaM} so as to distinguish explicitly  between the off-shell energy mode $\omega_k(M)=\sqrt{k^2+a^2 M^2}$  (formerly denoted just as $\omega_k$) and the on-shell one  $\omega_k(m)=\sqrt{k^2+a^2 m^2}$.  With this notation, lengthy but  straightforward calculations from equations  \eqref{EMTFluctuations} and \eqref{EMTRenormalized}  lead to the following compact result\cite{CristianJoan2020}:
\begin{equation}\label{Renormalized2}
\begin{split}
&\langle T_{00}^{\delta \phi}\rangle_{\rm Ren}(M)
=\frac{a^2}{128\pi^2 }\left(-M^4+4m^2M^2-3m^4+2m^4 \ln \frac{m^2}{M^2}\right)\\
&-\left(\xi-\frac{1}{6}\right)\frac{3 \mathcal{H}^2 }{16 \pi^2 }\left(m^2-M^2-m^2\ln \frac{m^2}{M^2} \right)+\left(\xi-\frac{1}{6}\right)^2 \frac{9\left(2  \mathcal{H}^{\prime \prime} \mathcal{H}- \mathcal{H}^{\prime 2}- 3  \mathcal{H}^{4}\right)}{16\pi^2 a^2}\ln \frac{m^2}{M^2}+\dots
\end{split}
\end{equation}
where dots stand just for higher adiabatic orders.
 The renormalized expression for the vacuum fluctuations, $\langle T_{\mu\nu}^{\delta \phi}\rangle_{\rm Ren}(M)$, is not yet the final one to extract the renormalized VED. As indicated in \eqref{EMTvacuum}, the latter is obtained from including the contribution from the $\rL$-term in the Einstein-Hilbert action \eqref{eq:EH}.   Therefore, the renormalized vacuum  EMT at the scale $M$ is given by
\begin{equation}\label{RenEMTvacuum}
\langle T_{\mu\nu}^{\rm vac}\rangle_{\rm Ren}(M)=-\rho_\Lambda (M) g_{\mu \nu}+\langle T_{\mu \nu}^{\delta \phi}\rangle_{\rm Ren}(M)\,.
\end{equation}
We distinguish between VED and ZPE: the latter is caused by the vacuum fluctuations of the fields, whereas the former combines the ZPE and the parameter $\rL$ in the action.
The renormalized VED  is precisely  the $00$th component of the above expression:
\begin{equation}\label{RenVDE}
\rho_{\rm vac}(M)= \frac{\langle T_{00}^{\rm vac}\rangle_{\rm Ren}(M)}{a^2}=\rho_\Lambda (M)+\frac{\langle T_{00}^{\delta \phi}\rangle_{\rm Ren}(M)}{a^2}\,,
\end{equation}
where we have used the fact that $g_{00}=-a^2$ in the conformal metric.    Explicitly, the VED comes out to be
\begin{equation}\label{RenVDEexplicit}
\begin{split}
\rv(M)&= \rho_\Lambda (M)+\frac{1}{128\pi^2 }\left(-M^4+4m^2M^2-3m^4+2m^4 \ln \frac{m^2}{M^2}\right)\\
&-\left(\xi-\frac{1}{6}\right)\frac{3 \mathcal{H}^2 }{16 \pi^2 a^2}\left(m^2-M^2-m^2\ln \frac{m^2}{M^2} \right)\\
&+\left(\xi-\frac{1}{6}\right)^2 \frac{9\left(2  \mathcal{H}^{\prime \prime} \mathcal{H}- \mathcal{H}^{\prime 2}- 3  \mathcal{H}^{4}\right)}{16\pi^2 a^4}\ln \frac{m^2}{M^2}+\cdots
\end{split}
\end{equation}
The (very)  interesting aspect about this form of renormalized VED is that the first two terms of this expression (i.e. those not depending on $\cH$) exactly cancel when we compute the difference of  $\rv$ values at two scales, say $M$ and $M_0$:
\begin{equation}\label{eq:VEDscalesMandM0Final}
\begin{split}
&\rv(M)-\rv(M_0)=\left(\xi-\frac16\right)\frac{3\cH^2}{16\pi^2 a^2}\,\left(M^2 - M_0^{2} -m^2\ln \frac{M^{2}}{M_0^2}\right)\\
&\phantom{XXXXXXXXx}+\left(\xi-\frac16\right)^2\frac{9}{16 \pi^2 a^4}\left(\mathcal{H}^{\prime 2}-2\mathcal{H}^{\prime \prime}\mathcal{H}+3 \mathcal{H}^4 \right)\ln \frac{M^2}{M_0^{2}}\,.\\
\end{split}
\end{equation}
To verify  the cancellation of the mentioned terms, we rewrite Einstein's equations\,\eqref{FieldEq2} using the renormalized parameters and including the higher derivative tensor  $H_{\mu \nu}^{(1)}$, which is necessary for renormalization purposes\cite{BirrellDavies82}. We need to write  only the vacuum part of the EMT since in doing the mentioned subtraction the background contribution of the field $\phi$  (and any other contribution, indicated below by \dots) will cancel, except the ($M$-dependent) change of the vacuum EMT at the two scales:
\begin{equation}\label{eq:EqsVac2}
\MPl^2 (M) G_{\mu \nu}+\alpha(M) H_{\mu \nu}^{(1)}= \langle T_{\mu\nu}^{\rm vac}\rangle_{\rm Ren}(M)+...
\end{equation}
On the \textit{r.h.s.} we have used Eq.\,\eqref{RenEMTvacuum}, which can be made more explicit using Eq,\,\eqref{Renormalized2}.  We may now subtract side by side  \eqref{eq:EqsVac2} at the scales $M$ and $M_0$ and project the $00$th component. Using the explicit form of $ G_{00}$ and $H_{00}^{(1)}$ in the FLRW metric we can perform the identifications on both sides of the subtracted equation. In particular, this renders specific expressions for the  shifts  $\MPl^2 (M)-\MPl^2 (M_0)$ and $\alpha(M)-\alpha(M_0)$, which we need not quote here\cite{CristianJoan2020, CristianJoan2021} . After performing these identifications, what is left of  $\langle T_{\mu\nu}^{\rm vac}\rangle_{\rm Ren}(M)- \langle T_{\mu\nu}^{\rm vac}\rangle_{\rm Ren}(M_0)$ must be zero. This is how we can prove that $\rL(M)-\rL(M_0)$ exactly cancels against the difference of the second term of \eqref{eq:VEDscalesMandM0Final} at the two scales (see Refs.\cite{CristianJoan2020, CristianJoan2021} for  more details):
\begin{equation}\label{eq:deltarL}
\begin{split}
&\left.\rL(M)\right|_{M_0}^M+\left.\frac{1}{128\pi^2 }\left(-M^4+4m^2M^2-3m^4+2m^4 \ln \frac{m^2}{M^2}\right)\right|_{M_0}^M\\
&=\rL(M)-\rL(M_0)+\frac{1}{128\pi^2}\left(-M^4+M_0^{4}+4m^2(M^2-M_0^{2})-2m^4\ln  \frac{M^{2}}{M_0^2}\right)=0\,.
\end{split}
\end{equation}
This important equality, enforced by the renormalized form of Einstein's equations, demonstrates that the renormalized EMT,  and in particular the renormalized VED, is free from quartic contributions associated to mass scales. That is why the relation \eqref{eq:VEDscalesMandM0Final} between the VED  at the two renormalization points $M$ and $M_0$ is a smooth function $\sim M^2\cH^2$.   With no quartic mass term surviving in this renormalization procedure, there is no need of fine tuning.   This is, of course, an extremely welcome feature, which obviously impinges on a possible solution of the CCP.
Another appealing (and novel) feature of our renormalization framework is the following: Eq.\,\eqref{eq:EqsVac2}  says that in Minkowski spacetime $\langle T_{00}^{\rm vac}\rangle_{\rm Ren}(M)=0$, and as a result the renormalized VED in flat spacetime is also zero in this context,  $\rv^{\rm Mink}=0$.

Let us now assume that we define the renormalized VED in the context of some Grand Unified Theory (GUT) scale  $M_0=M_X$, where typically  $M_X\sim 10^{16}$ GeV is associated with the inflationary scale.   We denote by $\rv(M_X)$  the value of the  VED at $M_X$.   We can relate $\rv(M_X)$  with the current value of the VED, $\rvo$,  assuming that  $\rv(M=H_0)=\rvo$,  where we choose  the second scale at  today's value of the Hubble parameter, $H_0$.  This quantity can be used as an estimate for the energy scale of the background gravitational field associated to the FLRW universe at present. We neglect  the $\sim H^4$ terms in the current universe.  Then, according to Eq.\,\eqref{eq:VEDscalesMandM0Final},   the connection between the two values of the  VED is
\begin{equation}\label{eq:rlMX1}
\rvo=\rv(M_X)+\frac{3}{16\pi^2}\left(\frac{1}{6}-\xi\right) H_0^2\left[M_X^2+{m^2}\ln \frac{H_0^{2}}{M_X^2}\right]\,,
\end{equation}
or swapping terms on both sides:
\begin{equation}\label{eq:rlMX2}
\rv(M_X)=\rvo-\frac{3\nu}{8\pi}\,H_0^2\,\mpl^2\,.
\end{equation}
Here we have defined the `running  parameter' for the VED:
\begin{equation}\label{eq:nueff}
\nu\equiv\frac{1}{2\pi}\,\left(\frac{1}{6}-\xi\right)\,\frac{M_X^2}{\mpl^2}\left(1+\frac{m^2}{M_X^2}\ln \frac{H_0^{2}}{M_X^2}\right)\,.
\end{equation}
We naturally expect $|\nu|\ll1$,  owing to the ratio $M_X^2/\mpl^2\ll1$.   The vanishing of $\nu$ and hence of  the dynamical $\sim H^2$ part of \eqref{eq:rlMX1} is obtained only for conformal coupling: $\xi=1/6$.
There are, however,  fermionic contributions to $\nu$ as well. They do not depend on $\xi$, of course, but shall not be addressed  here\cite{JCS2021}. The accurate determination of  $\nu$  can only be obtained by fitting the RVM to the overall cosmological data, as it has been done  e.g. in \cite{ApJL2015,RVMphenoOlder1,RVMpheno1,RVMpheno2,Mehdi2019,EPL2021}. These analyses show that $\nu$ is of order $10^{-3}$ and positive \footnote{It is interesting to remark that even Brans-Dicke gravity with a cosmological constant can mimic the RVM, as shown in \cite{GRF2018,JavierJoan2018}, with a value of $\nu$ of the same order\,\cite{BD2020}. See also \cite{SinghJoan2021}.}.
From \eqref{eq:rlMX2} and \eqref{eq:VEDscalesMandM0Final}  we can approximately estimate the VED near our time by  taking  $M$ of order of the energy scale defined by the numerical value of $H$ around the current epoch\cite{CristianJoan2020}:
\begin{equation}\label{eq:RVM2}
\rv(H)\simeq \rvo+\frac{3\nu}{8\pi}\,(H^2-H_0^2)\,\mpl^2=\rvo+\frac{3\nu}{8\pi G_N}\,(H^2-H_0^2)\,.
\end{equation}
This expression reproduces the canonical form of the  RVM\cite{JSPRev2013}.  A generalized form of the RVM is possible in which an additional term proportional to $\dot{H}$ (of order $H^2$) can be included on the \textit{r-h.s.} of \eqref{eq:RVM2} with another small coefficient $\tilde{\nu}$, see\cite{CristianJoan2020,CristianJoan2021} for details.  As noted, we have neglected the ${\cal O}(H^4)$ terms to derive the previous formulas. These terms can have an  impact only  for the early universe and produce inflation, but we refer once more the reader to \cite{CristianJoan2020,CristianJoan2021}  for an expanded  discussion.

\section{Phenomenological applications: $H_0$ and $\sigma_8$ tensions}\label{sec:PhenoApplications}

We are now ready to produce some phenomenological output using the low-energy RVM form \eqref{eq:RVM2}, or even the generalized one containing the $\sim \dot{H}$ term. In fact, we can use the following extended RVM structure for the VED\cite{CristianJoan2020,CristianJoan2021}:
\begin{equation}\label{eq:RVMvacuumdadensity}
\rv(H) = \frac{3}{8\pi{G}_N}\left(c_{0} + \nu{H^2+\tilde{\nu}\dot{H}}\right)+{\cal O}(H^4)\,.
\end{equation}
The additive constant $c_0$ is fixed by the boundary condition $\rho_{\rm vac}(H_0)=\rvo$.  For recent expositions of the RVM and its many phenomenological applications making use of the above form of VED, see \cite{AdriaPhD2017,JavierPhD2021} and references therein\footnote{It is interesting to mention that at the  pure cosmographic/cosmokinetic level (hence in a more model-independent way)   the RVM appears also as a favoured  model. For example, the form \eqref{eq:RVMvacuumdadensity}  has been used in \cite{Mehdi2021} to study the different types of DE models  in the framework of the cosmographic approach. Using the Hubble diagrams for  SnIa, quasars, gamma-ray bursts  as well as the data on baryonic acoustic oscillations in different combinations, it is found that the RVM  fits better the cosmographic data than other DE models, including the concordance $\CC$CDM.}.
In what follows I will summarize the findings of the recent study\cite{EPL2021}, which involves the most complete set of cosmological data to date\footnote{See \cite{JSPRev2016} for a review of previous analyses  repeatedly emphasizing the possible impact of the RVM phenomenology on improving the description of the observations as compared to the $\CC$CDM.}.  Since the parameters $\nu$ and $\tilde{\nu}$ are ultimately to be fitted to observations, we shall simplify our analysis here assuming one single parameter with the choice $\tilde{\nu}=\nu/2$.  This does not change anything fundamental.  As a result, if we neglect the higher order terms, Eq.\,\eqref{eq:RVMvacuumdadensity} takes on the suggestive  form
\begin{equation}\label{RRVM}
\rv(H) =\frac{3}{8\pi{G_N}}\left(c_0 + \frac{\nu}{12} {R}\right)\equiv \rv(R)\,,
\end{equation}
where  ${R} = 12H^2 + 6\dot{H}$ is the curvature scalar. For this reason we may call this form of the VED the  `RRVM'.
Such a RRVM implementation has the double advantage of using one single parameter and provides  a safe path to the early epochs of the cosmological evolution since when we approach the radiation dominated era we have $R\simeq 0$, or to be more precise:  $R/H^2\ll 1$. This fact insures that no conflict is generated with the BBN constraints.  Early on the RVM has its own mechanism for inflation (as we have already mentioned), but we have no room to address these aspects here, see\cite{LimaBasSol,JSPRev2015,GRF2015,Yu2020}.  With one and the same VED given by Eq.\,\eqref{RRVM} we may consider two types of RRVM scenarios, to wit:  type I scenario, in which the vacuum interacts with matter; and  type II,  where matter is conserved at the expense of an exchange between the vacuum and a mildly evolving gravitational coupling $G (H)$.   For type I models we assume that  the vacuum exchanges energy with cold dark matter (CDM) only, as follows:
\begin{equation}\label{eq:LocalConsLaw}
\dot{\rho}_{dm} + 3H\rho_{dm} = -\dot{\rho}_{\rm vac}\,.
\end{equation}
Solving for the matter densities one finds\cite{EPL2021}
\begin{equation}\label{eq:MassDensities}
\rho_m(a) = \rho^0_m{a^{-3\xi}}\,, \ \ \ \rho_{dm}(a) = \rho^{0}_m{a^{-3\xi}}  - \rho^0_b{a^{-3}} \,,
\end{equation}
where $\xi \equiv \frac{1 -\nu}{1 - \frac{3}{4}\nu}$.
They recover the $\CC$CDM form for $\xi=1$ ($\nu=0$).  The small departure of $\nu$ from zero (or $\xi$ from one)  is what permits a mild dynamical vacuum evolution:
\begin{align}  \label{Vacdensity}
\rv(a) &= \rvo + \left(\frac{1}{\xi} -1\right)\rho^0_m\left(a^{-3\xi} -1\right)\,.
\end{align}
For type I models we  admit also the  possibility that the dynamics of vacuum is relatively recent (see e.g. \cite{Martinelli2019}).  For instance, one may assume that the vacuum became dynamical in the form \eqref{Vacdensity} at a threshold redshift $z_{*}\simeq  1$ (so $\rv(z) = \rvo$ for $z>z_{*}$).   We shall compare this option with the situation when there is no such threshold.  As for type II models,  matter is conserved (hence no exchange with vacuum), but the vacuum can still evolve as long as the effective gravitational coupling also evolves (very mildly) with the expansion $G_{\rm eff}=G_{\rm eff}(H)$, starting from an initial value (which enters our fit). In this case, we do not consider the effect of the threshold because it proves to be much smaller.  One can show that the approximate behavior of the VED in the present time is (recall that $|\nu|\ll1$)\cite{EPL2021}:
\begin{equation}\label{eq:VDEm}
\rv(a)=\frac{3c_0}{8\pi G_N}(1+4\nu)+\nu\rho_m^{0}a^{-3}+\mathcal{O}(\nu^2)\,.
\end{equation}
Once more, for $\nu=0$ the VED remains constant at the value $\rv={3c_0}/(8\pi G_N)=\CC/(8\pi G_N)$, but otherwise it shows a moderate dynamics as in the type I case.   One can also show that the effective gravitational coupling evolves approximately as  $G_{\rm eff}(a)\simeq G_N (1+\epsilon\ln\,a)$ in the current epoch (with $0<\epsilon\ll 1$ of order $\nu$), thus confirming the very mild (logarithmic) evolution of $G$. Since $\epsilon>0$ the gravitational strength exhibits an asymptotically free behavior (i.e. being smaller in the past).

\begin{figure}[!t]
\begin{center}
\includegraphics[width=4.3in]{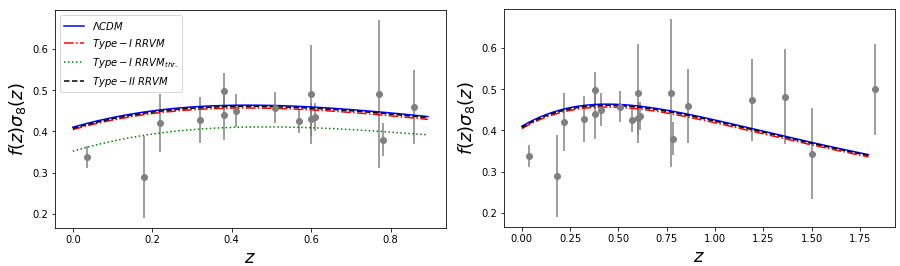}
\end{center}
\caption{Theoretically predicted curves of $f(z)\sigma_8 (z)$ for the various models (type I and type II), testing the possible existence of a  threshold redshift  in the case of type I.   The data points employed in our analysis are described in \cite{EPL2021}. We present the results in two different redshift windows.  It can be seen that the type I RRVM with threshold redshift $z_{*}\simeq 1$  has a most visible and favorable impact on solving the $\sigma_8$ tension.}
\vspace{-0.3cm}
\protect\label{fig:sugra}
\end{figure}

For an accurate comparison of the theoretical predictions and the observations,   the  LSS  formation data are also of paramount importance, all the more if we take into account that one of the aforementioned $\CC$CDM tensions (the $\sigma_8$ one) stems from it.  As we shall see, allowing for some evolution of the vacuum can be the clue to alleviate the $\sigma_8$ tension since such dynamics affects significantly the cosmological perturbations\cite{RVMsigma8}.
We consider the perturbed, spatially flat,  FLRW metric   $ds^2=-dt^2+(\delta_{ij}+h_{ij})dx^idx^j$, in which $h_{ij}$ stands for the metric fluctuations. These fluctuations are nontrivially coupled to the  matter density perturbations $\delta_m=\delta\rho_m/\rho_m$.
We have implemented the full perturbations analysis  in the context of the Einstein-Boltzmann code \texttt{CLASS}\cite{CLASS}  (in the synchronous gauge)\cite{EPL2021}.   Since baryons do not interact with the time-evolving VED the perturbed conservation equations are not directly affected. However, the corresponding equation for CDM is modified in the following way:
\begin{equation}\label{eq:perturbCDM}
\dot{\delta}_{dm}+\frac{\dot{h}}{2}-\frac{\dot{\rho}_{\rm vac}}{\rho_{dm}}\delta_{dm}=0\,,
\end{equation}
with $h=h_{ii}$ denoting the trace of $h_{ij}$. We remark that  the term $\dot{\rho}_{\rm vac}$ is nonvanishing for these models. Thus, it affects the fluctuations of CDM in a way which produces a departure from the $\CC$CDM. The above equation is, of course, coupled with the metric fluctuations and the combined system must be solved numerically.
The analysis of the linear LSS regime is performed with the help of the weighted linear growth $f(z)\sigma_8(z)$, where $f(z)$ is the growth factor and $\sigma_8(z)$ is the rms mass fluctuation amplitude on scales of $R_8=8\,h^{-1}$ Mpc at redshift $z$.   The quantity $\sigma_8(z)$ is directly provided by \texttt{CLASS}. Similarly,  we can extract the  (observationally measured) linear growth function  $f(a)$ directly from the matter power spectrum $P_m(a,\vec{k})$, which is computed numerically by \texttt{CLASS} under adiabatic
initial conditions.  The results  can be seen in Figures 1 and 2 for the various models.

\begin{figure}[!t]
\begin{center}
\includegraphics[width=4.3in]{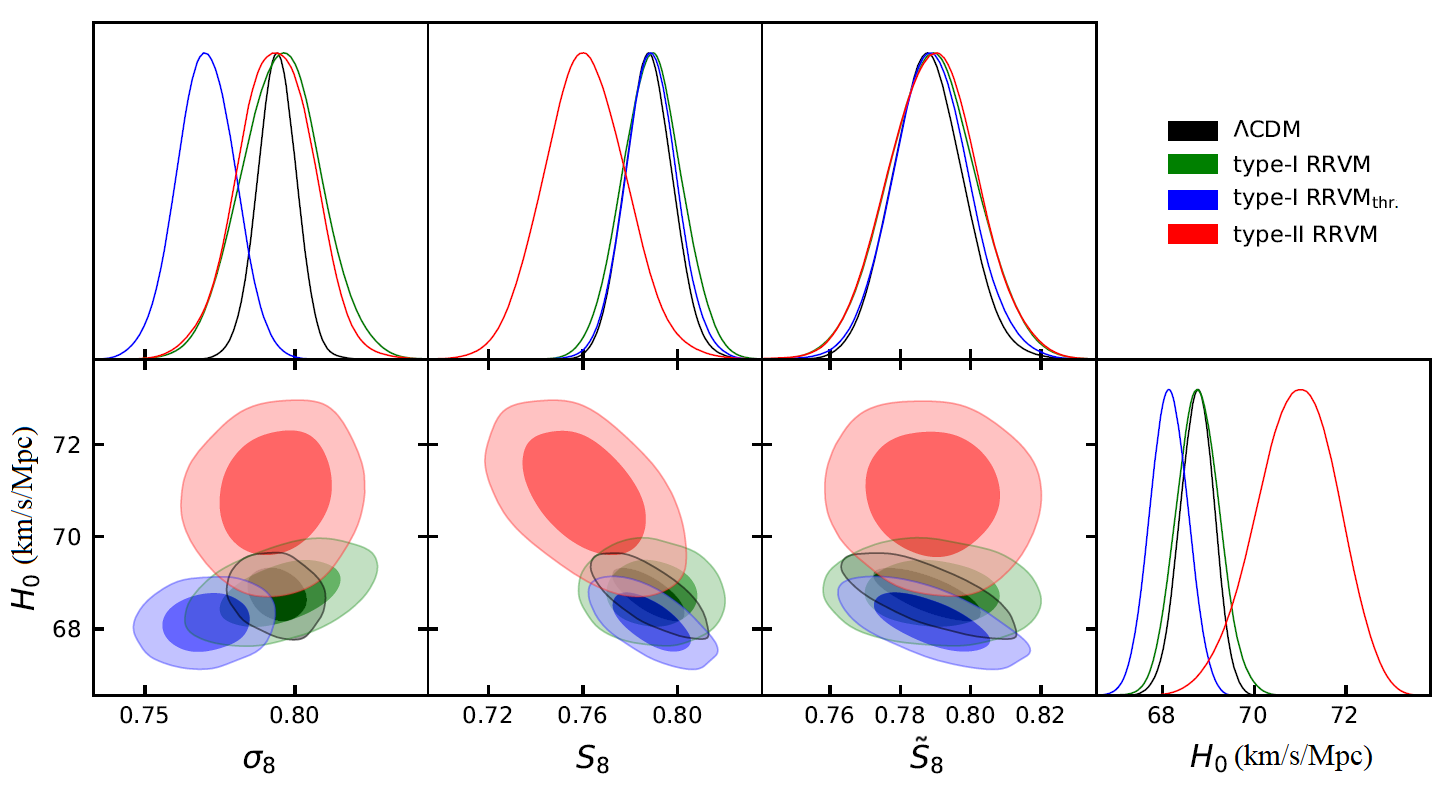}
\end{center}
\caption{ The $1\sigma$ and $2\sigma$ c.l.  contours in the $H_0$-$\sigma_8,S_8,\tilde{S}_8$ planes and the corresponding one-dimensional posteriors for the  RRVM's obtained from the  Baseline+$H_0$ data set (meaning that the local value of $H_0$ is included along with the remaining data, see \cite{EPL2021}).  It is apparent that the  type II model alleviates the $H_0$ tension without spoiling the $\sigma_8$ one, whereas the type I model with threshold redshift $z_{*}\simeq 1$ can fully fix the latter (see also Fig.\,1) but cannot reduce the former.}
\vspace{-0.3cm}
\protect\label{fig:sugra}
\end{figure}

To compare the  RRVM's (types I and II) with the $\CC$CDM, we have defined a joint likelihood function ${\cal L}$. The overall fitting results are reported in Tables 1 and 2. The total $\chi^2$ to be minimized in our case is given by
\be
\chi^2_{\rm tot}=\chi^2_{\rm SnIa}+\chi^2_{\rm BAO}+\chi^2_{ H}+\chi^2_{\rm f\sigma_8}+\chi^2_{\rm CMB}\,.
\ee
The above $\chi^2$ terms are defined in the standard way from the data including the covariance matrices\cite{DEBook}.  In particular, the $\chi^2_{H}$ part may contain or not the local $H_0$ value measured by Riess, see \cite{EPL2021}. The local determination of $H_0$ (which is around $4\sigma$ away from the corresponding Planck 2018 value based on the CMB) is the origin of the so-called $H_0$ tension\cite{tensions,PerivoSkara2021,ValentinoReviewTensions,TensionsJSP2018}.
Taking into account that the RRVM's of type I and II have one and two more parameters, respectively, as compared to the $\CC$CDM, a fairer  model comparison is achieved by computing the numerical differences between the Deviance Information Criterion (DIC) of the $\CC$CDM model against the RRVM's: $\Delta{\rm DIC}={\rm DIC}_{\rm \CC CDM}-{\rm DIC}_{\rm RRVM}$.  These differences will be (and in fact are) positive if the RRVM's fit better the overall data than the $\CC$CDM. The DIC is defined by\cite{DIC}
$  {\rm DIC}=\chi^2(\overline{\theta})+2p_D\,.$
Here $p_D=\overline{\chi^2}-\chi^2(\overline{\theta})$ is the effective number of parameters of the model, and $\overline{\chi^2}$ and $\overline{\theta}$ the mean of the overall $\chi^2$ distribution and of the parameters, respectively.  The posterior distributions and corresponding constraints for the various dataset combinations have been obtained with \texttt{Montepython}\cite{Montepython} in combination with \texttt{CLASS}\cite{CLASS}.

The DIC value can be obtained  from the Markov chains generated with \texttt{MontePython}.
If  $+5<\Delta{\rm DIC} <+10$  one should conclude strong evidence favoring the RRVM's over the $\CC$CDM. For $\Delta{\rm DIC}>+10$  it is said that the evidence is very strong. Such is the case when we use a threshold  redshift $z_{*}\simeq 1$ in type I RRVM.  In stark contrast, when the threshold is removed we find only moderate evidence against it (viz. $-3<\Delta{\rm DIC} <-2$), although the fitting performance is still slightly better (smaller $\chi^2_{\rm min}$) than the $\Lambda$CDM.  The effect of the threshold can be very important and suggests that a mild dynamics of the vacuum is welcome, especially if it gets activated at around the  epoch when the vacuum dominance appears, namely at $z\simeq 1$.  Unfortunately, type I models with fixed $G_{\rm eff}=G_N$ do not help an inch to solve the $H_0$ tension since the value of $H_0$ remains stuck around the CMB value\cite{EPL2021}.  In stark contrast, type II models can alleviate the two tensions at a time. The overall $\Delta{\rm DIC}$ value of the fit is quite significant ($+5.5$), still in the strong evidence region, providing values of $H_0$ markedly higher as compared to type I models (specifically $H_0=70.93^{+0.93}_{-0.87}$ Km/s/Mpc\cite{EPL2021}) along with $\sigma_8$ and $S_8$ values in the needed moderate range ($\sigma_8=0.794^{+0.013}_{-0.012}$ and $S_8=0.761^{+0.018}_{-0.017}$)\cite{EPL2021}. The values of $S_8$ in all RRVM's are perfectly compatible with recent weak lensing and galaxy clustering measurements\cite{Heymans2020}. For type II models a related observable analogous to (but different from) $S_8$ is $\tilde{S}_8=S_8\sqrt{G_{\rm eff}(0)/G_N}$ .  We show the corresponding contours in Fig. 2. The net outcome of this analysis is that  the only model capable of alleviating the two tensions  ($H_0$ and $\sigma_8)$ is RRVM of type II, whereas the type I model can (fully) solve the $\sigma_8$ tension but has no bearing on the $H_0$ one.


\section{Conclusions and outlook} \label{sec:conclusions}

 In this short review presentation I have described  the renormalization of the energy-momentum tensor (EMT) of a real quantum scalar field non-minimally coupled to classical gravity in the cosmological context.  The main aim was to show that, out of the very quantum effects of matter, there emerges  the running vacuum model (RVM) structure, which is  the effective form of the renormalized vacuum energy density (VED). The method is based on an off-shell extension of the adiabatic regularization and  renormalization procedure, which we have used in the recent works\cite{CristianJoan2020,CristianJoan2021} and where for the first time we provided a calculation of the zero point energy (ZPE) of a quantum scalar field that is free from the need of extreme fine tuning. The latter is well-known to be one of the most striking and  bizarre aspects of the  cosmological constant problem\cite{Weinberg89}. The absence of need for  fine-tuning is related to  the non appearance in our framework of the terms which are proportional to the quartic mass of the fields,  i.e.  $\sim m^4$.  In the standard model of particle physics, these terms  are usually responsible for the exceedingly large contributions to the VED and requires preposterous fine-tuning with the renormalized vacuum parameter $\rL$ in the action\,\cite{JSPRev2013}.  The calculational procedure in our approach is based on the WKB expansion of the field modes in the  FLRW spacetime and  the use of an appropriately renormalized EMT. The latter is obtained by performing a substraction  of its on-shell value (i.e. the value defined  on the mass shell  $m$ of the quantized field)  at an arbitrary renormalization point $M$. The resulting EMT becomes finite because we subtract  the first four adiabatic orders (the only ones that can be divergent).  Since the off-shell renormalized EMT  is a function of  the arbitrary renormalization point $M$, we can compare the renormalized result at different epochs of the cosmic history.  The introduction of such `sliding' scale leads to the renormalization group (RG) flow, which in the case of FLRW cosmology we may associate  in a natural way  with Hubble's expansion rate $H$.

 While the RG approach was actually  the first qualitative idea behind  the RVM\,\cite{JSPRev2013}, with the present QFT calculations in curved spacetime we provide for the first time\cite{CristianJoan2020,CristianJoan2021}  a solid foundation of the  RVM, in which the dynamical structure of the VED is seen to emerge from the quantum effects associated with the adiabatic  renormalization of the EMT.   Let us  mention that even though our QFT calculation has been simplified by the use of a single (real) quantum scalar field, further investigations show that the generalization of  these results for  multiple fields,  involving scalar as well as  vector and fermionic components, lead as well to the generic RVM structure mentioned here up to (nontrivial) computation details\cite{JCS2021}.

 At the end of the day, we have been able to show that the genuine form of the VED for the current universe can be achieved from direct calculations of QFT in the FLRW spacetime.   In such structure, the powers of $H$ (and its time derivatives)  are of even adiabatic order.  This means that all of the allowed powers  effectively carry an even number of time derivatives of the scale factor, which is essential to preserve the general covariance of the action.  Linear terms (cubic, or in general of odd order) in $H$ are incompatible with such a covariance and in fact do not appear in the final result.  To be more precise, all terms with an odd number of time derivatives of the scale factor are ruled out.  The form of  $\rho_{\rm vac}(H)$ predicted by the RVM  at low energies is remarkably simple but it certainly goes beyond a rigid cosmological constant term.  It consists of an additive constant  (to be identified basically with the cosmological term)  together with a small dynamical component  $\sim \nu H^2$, in which the dimensionless parameter $\nu$ can be computed from the underlying QFT framework and is predicted to be small  ($|\nu|\ll1$).  Ultimately, its value can only be known  upon fitting the RVM to the overall cosmological data.  The physical  outcome is that today's cosmic vacuum  is mildly dynamical. In fact, in previous works the model has been phenomenologically fitted to a large wealth of cosmological data and the running parameter $\nu$ has been found to be positive and in the ballpark of $\sim 10^{-3}$\, cf.\cite{ApJL2015,RVMphenoOlder1,RVMpheno1,RVMpheno2}.

Recent phenomenological analyses involving a large set of updated cosmological data on SnIa+H(z)+BAO+LSS+CMB\,\cite{EPL2021,CosmoTeam2021} and analyzed with the Boltzmann code \texttt{CLASS}\cite{CLASS}  confirm to a large extent the results reported in the aforementioned works, which were carried out within an approximate treatment of the CMB. The basic result is still the same, they point to substantial evidence that a  mild dynamics of the cosmic vacuum is  helpful to describe the overall cosmological observations as compared to the standard cosmological model with  a  rigid $\CC$-term.  From our analysis of two variants of the RVM, we have found that  for type I models the $\sigma_8$ tension can be fully overcome  ($\lesssim 0.4\sigma$ c.l.)  provided there exists a threshold  redshift $z_{*}\simeq 1$ where the vacuum dynamics is triggered. Solving the $H_0$ tension, however, proves more demanding as it requires the combination of vacuum dynamics with running $G$, which is the characteristic of the type II models.  Interestingly enough, the two tensions can actually  be dealt with at a  time, the $H_0$ remaining at $\sim 1.6\sigma$  and the $\sigma_8$  one at $\sim 1.3\sigma$ (or at only $\sim 0.4\sigma$ if stated in terms of $S_8$)  \cite{Heymans2020}. The successful cutback of the two tensions is highly remarkable and is strongly supported by standard information criteria, such as the deviance information criterion (DIC).  More work will be needed, of course, to confirm if the RVM can fully solve the $\sigma_8$ and $H_0$ tensions\cite{CosmoTeam2021}. It  will depend also on  the upcoming data in the next few years.

\vspace{-0.1cm}
\section*{Acknowledgements}
It is my pleasure to thank  N.E. Mavromatos for  inviting me to speak in the DM1 parallel session: ``Interacting Dark Matter'' of the MG16 Marcel Grossmann virtual Conference, July 5-10 2021.   Work  partially funded by  projects  PID2019-105614GB-C21 and FPA2016-76005-C2-1-P (MINECO, Spain), 2017-SGR-929 (Generalitat de Catalunya) and CEX2019-000918-M (ICCUB).  The author acknowledges participation in the COST Association Action CA18108 ``{\it Quantum Gravity Phenomenology in the Multimessenger Approach (QG-MM)}''. This presentation  is  based in part on works with A.  G\' omez-Valent,  J. de Cruz P\'erez and C. Moreno-Pulido. I thank them warmly for the enjoyable collaboration.

\end{document}